\documentclass[aps,prl,twocolumn,superscriptaddress,showpacs,floatfix]{revtex4}
\usepackage{graphicx}
\usepackage{epsfig}
\usepackage{amsmath}
\usepackage{amssymb}

\begin{document}


\title{MnO Spin-Wave Dispersion Curves from Neutron Powder Diffraction Data}

\author{Andrew L. Goodwin}
\affiliation{Department of Earth Sciences, Cambridge University,
Downing Street, Cambridge CB2 3EQ, U.K.}

\author{Martin T. Dove}
\affiliation{Department of Earth Sciences, Cambridge University, Downing Street, Cambridge CB2 3EQ, U.K.}

\author{Matthew G. Tucker}
\affiliation{ISIS Facility, Rutherford Appleton Laboratory, Chilton,
Didcot, Oxfordshire OX11 0QX, U.K.}

\author{David A. Keen}
\affiliation{ISIS Facility, Rutherford Appleton Laboratory, Chilton,
Didcot, Oxfordshire OX11 0QX, U.K.} \affiliation{Department of Physics,
Oxford University, Clarendon Laboratory, Parks Road, Oxford OX1 3PU,
U.K.}

\date{\today}
\begin{abstract}
We describe a model-independent approach for the extraction of
spin-wave dispersion curves from neutron total scattering data. The
method utilises a statistical analysis of real-space spin
configurations to calculate spin-dynamical quantities. The {\sc
rmcp}rofile implementation of the reverse Monte Carlo refinement
process is used to generate a large ensemble of supercell spin
configurations from powder diffraction data. Our analysis of these
configurations gives spin-wave dispersion curves that agree well with
those determined independently using neutron triple-axis spectroscopic
techniques.
\end{abstract}

\pacs{75.30.Ds, 61.12.Bt, 02.70.Uu}

\maketitle

In this Letter, we explore the unexpected possibility that neutron
powder diffraction might be used to probe spin \emph{dynamics} in
magnetic materials. In effect, what we are asking is whether there
is sufficient information preserved in experimentally observed
neutron scattering functions to allow reconstruction of the
spin-wave dispersion.

It is well known that the intensities of magnetic Bragg reflections
are determined entirely by one-particle correlations (the same is
true, naturally, for the nuclear Bragg reflections). In terms of
spin-dynamical information, these describe the average fluctuations
in moment at each magnetic site. However, any determination of
spin-wave dispersion relations would require access to the
two-particle spin correlation functions. For this reason, we have
focussed our efforts on an analysis of the \emph{total} magnetic
neutron scattering patterns, considering both elastic (which
includes Bragg) and inelastic (included in the diffuse) components
alike~\cite{Dove_2002,totalscattering}. This is because the latter
cannot be described effectively without recourse to the two-particle
correlation functions; indeed, all the spin-dynamical information is
folded into the total scattering process. What is not clear is the
extent to which the orientational averaging implicit in powder
diffraction experiments complicates---or indeed precludes---recovery
of this information. Our paper aims to address the two issues so
raised: (i) to what extent are the two-particle spin correlations
preserved in total neutron scattering data, and (ii) how might one
extract these correlations in practice, and in doing so determine
spin-wave dispersion relations? The potential impact of a robust
methodology based on neutron powder diffraction is significant,
particularly since spin-wave information could be rapidly obtained
from \emph{e.g.} newly-discovered polycrystalline materials and/or
those materials for which a route to large single crystal samples
does not exist.

A similar problem---one that has received more widespread
attention---is that of extracting phonon dispersion curves from the
neutron powder diffraction patterns of non-magnetic
materials~\cite{phonons,Goodwin_2004,Goodwin_2005}. On the one hand,
the extraction of spin-waves might be expected to prove more
tractable, in that the number of spin-wave modes in a material is
always less than the number of phonon modes, and the absolute
energies of spin-waves tend to be lower than most phonon energies
(and hence their signature within diffraction data should be
clearer). On the other hand, the analysis of diffraction data from
magnetic materials will be complicated both by the superposition of
nuclear and magnetic scattering contributions and by the magnetic
form-factor dependence. What is clear---and this has emerged in the
development of methods for extracting phonons from diffraction---is
the need for model-independent approaches, so that the answers one
obtains are driven by data rather than the choice of any particular
dynamical model~\cite{Goodwin_2004,Goodwin_2005}.

This paper describes a ``proof of principle'' investigation into the
possibility of extracting spin-wave dispersion curves from powder
diffraction data. We present results for manganese(II) oxide (MnO)
that indicate one can indeed recover the spin-wave excitation
spectrum for this system, albeit with some identifiable yet
surmountable limitations. Our paper begins by describing the reverse
Monte Carlo (RMC) method of refining total neutron scattering data.
We proceed to illustrate how statistical analysis of large ensembles
of these RMC configurations can yield quantitative measurements of
the spin-wave dispersion. Our diffraction-based MnO spin-wave
dispersion curves are presented and compared with those obtained
independently using triple-axis inelastic neutron spectroscopy. We
close with a discussion of some necessary developments for the
approach to offer a viable method of measuring spin-wave dispersion
curves in materials for which established techniques are
impractical.

For non-magnetic materials, the Fourier transform of the observed
total scattering data gives the real-space atomic pair distribution
function, which contains detailed information regarding the local
structural environments present in the sample. The RMC method
provides a particularly effective means of analysing total
scattering data, in that it can be used to refine atomistic
configurations that represent large supercells of the known
structural unit cell~\cite{McGreevy_2001,Keen_2005}. These
supercells reflect simultaneously the average periodic structure (as
given by the observed Bragg intensities) together with the
collective displacements from which the diffuse scattering arises.
We have shown elsewhere that these atomic displacements are well
described as an instantaneous superposition of all phonon
modes~\cite{Goodwin_2004,Goodwin_2005}.

Magnetic materials require refinement of both atomistic and spin
configurations to account for the observed scattering function. We have
recently extended the {\sc rmcp}rofile implementation of the RMC method
to allow simultaneous modelling of nuclear and magnetic
structures~\cite{Goodwin_2006}. The procedure, which is described in
detail elsewhere~\cite{Goodwin_2006,Dove_2002}, involves minimisation
of the ``mismatch'' function
\begin{equation}
\chi^2_{\rm RMC}=\chi^2_{S(Q)}+\chi^2_{\rm Bragg},
\end{equation}
where
\begin{eqnarray}
\chi^2_{S(Q)}&=&\sum_k\sum_j[S_{\rm calc}(Q_j)_k-S_{\rm
exp}(Q_j)_k]^2/\sigma^2_k(Q_j),\nonumber\\
\chi^2_{\rm Bragg}&=&\sum_i[I_{\rm calc}(t_i)-I_{\rm
exp}(t_i)]^2/\sigma^2(t_i).
\end{eqnarray}
Here, $S(Q)$ is the observed scattering function as measured in each
of $k$ data sets (such as from a range of detector banks, for
example), and $I(t)$ is the time-of-flight Bragg powder profile as
described in more detail elsewhere~\cite{Keen_2005}. The function
$\chi^2_{\rm RMC}$ is minimised by successive random moves of two
sorts: displacement of atoms in the atomistic configuration and
re-orientation of individual spin vectors in the associated spin
configuration. Such moves are continually generated and accepted or
rejected according to the Monte Carlo algorithm until the
fit-to-data does not improve further. At this point, one has a pair
of configurations (one atomistic, one spin) that reflects both
nuclear and magnetic contributions to the Bragg \emph{and} diffuse
scattering patterns. As such, we can expect the RMC spin
orientations to represent an instantaneous superposition of the
spin-wave modes: a ``snapshot'' of the spin dynamics. The essence of
our analysis is to collect large numbers of these spin
configurations, so as to sample many different ``snapshots'' of the
spin-dynamical motion. By probing the various spin correlations that
recur, and the frequency with which they do so, we will show how one
can reconstruct spin-wave dispersion curves from such ensembles.

We have elected to use MnO as a case-study in this investigation as
its magnetic structure and spin dynamics have been the focus of a
large body of systematic
investigation~\cite{mno,Goodwin_2006,Pepy_1974,Kohgi_1974,Collins_1973}.
It is widely considered a benchmark antiferromagnetic material, and
is commonly used as a representative of the family of first-row
transitional-metal oxides in a variety of theoretical studies. A
paramagnet at room temperature, MnO exhibits long-range
antiferromagnetic ordering at temperatures below $T_{\rm N}=118\,$K.
Its paramagnetic to antiferromagnetic transition is accompanied by a
distortion of the high-temperature cubic lattice to a monoclinic
variant with pseudo-rhombohedral geometry. The basic
antiferromagnetic structure of MnO can be described in terms of a
single spin-alignment axis, and its spin dynamics are
well-understood in terms of simple Heisenberg interactions.

We collected neutron total scattering data for a powdered sample of
MnO using the GEM instrument at ISIS~\cite{Hannon_2005}. The
experiment spanned a range of momentum transfers $0.3 < Q <
50\,$\AA$^{-1}$ and was performed at a temperature of 100\,K. We
expected the increased relative magnitude of spin displacements at
this temperature to facilitate our analysis: a study at more
``conventional'' temperatures (\emph{e.g.}, where the spin
excitation energies are less broad) would place a greater emphasis
on resolution effects than proof of concept. Raw data were converted
to $S(Q)$ and Bragg intensity data for use as input for the RMC
procedure. We used the {\sc rmcp}rofile program to refine an
ensemble of \emph{ca} 600 RMC configurations, each of which
contained a $20\times20\times20$ array of fcc unit cells
(\emph{i.e.}, a $10\times10\times10$ array of conventional magnetic
unit cells, containing 32\,000 Mn atoms). Our starting
configurations were prepared by assigning to each atom a small
random displacement from its average position; the spin orientations
were also varied in a similar fashion. For consistency with previous
spin-wave investigations \cite{Pepy_1974,Kohgi_1974,Collins_1973},
we neglected any deviations from cubic lattice symmetry; this
simplification is known to have little effect on the associated
analysis. Each pair of equilibrium configurations were separated by
a minimum of 150\,000 RMC ``moves''~\cite{separation_note}, so that
they might be considered essentially independent for our analysis.

A method of extracting spin-dynamical information from ensembles of
spin configurations will be published in detail
elsewhere~\cite{Goodwin_preprint}; there is a strong analogy to the
known methods of extracting lattice-dynamical information from
atomistic configurations~\cite{phonontheory}. Here we describe the
most pertinent aspects of the theory. The analysis begins by
calculating for each configuration (labelled $t$) a number of
collective variables of the form
\begin{equation}
\tau(j,\mathbf k,t)=\sqrt{\frac{\hbar
S_j}{2N}}\sum_\ell\sigma^+(j\ell,t)\exp[{\rm i}\mathbf k\cdot\mathbf
r(j\ell)].
\end{equation}
There will be one of these variables for each of the $Z$ spins in
the (primitive) magnetic unit cell. Here, $S_j$ is the spin quantum
number of the spin type $j$, $\mathbf r(j\ell)$ its average position
in unit cell $\ell$, and $\sigma^+(j\ell,t)$ a parameter that
describes the deviation from the average spin alignment axis. All
$Z$ collective variables at each wave-vector $\mathbf k$ are
assembled into the one column vector $\boldsymbol\tau(\mathbf k,t)$.
This quantity is related to the Holstein-Primakoff magnon variable
$\mathbf b_{\mathbf k}(t)$~\cite{Holstein_1940} via the mapping
\begin{equation}\label{cob}
\sqrt\hbar\mathbf b_{\mathbf k}(t)=\mathbf A(\mathbf
k)\cdot\boldsymbol\tau(\mathbf k,t).
\end{equation}
Here, the change of basis occurs between the normal mode coordinates
(the basis of $\mathbf b$) and the spin-type coordinates (the basis of
$\boldsymbol\tau$), and is given by the ($t$-independent) matrix
$\mathbf A$, itself constructed from the spin-wave mode displacement
vectors~\cite{Goodwin_preprint}.

We proceed by calculating the $t$-averaged matrix $\boldsymbol\Sigma$,
\begin{equation}\label{sigma}
\boldsymbol\Sigma(\mathbf k)=\langle\boldsymbol\tau^\ast(\mathbf
k)\boldsymbol\tau^{\rm T}(\mathbf k)\rangle.
\end{equation}
This representation is useful because it allows us to exploit the
orthonormality of the basis for $\mathbf b$: in particular, the
matrix $\mathbf b_{\mathbf k}^\ast(t)\mathbf b_{\mathbf k}^{\rm
T}(t)$ is diagonal with entries given by the mode occupation numbers
$n(\mathbf k)$~\cite{Holstein_1940}. Substitution of Eq.~\eqref{cob}
into Eq.~\eqref{sigma} gives
\begin{equation}
\mathbf A(\mathbf k)\cdot\boldsymbol\Sigma(\mathbf k)\cdot\mathbf
A^{\rm T}(\mathbf k)=\hbar\langle\mathbf b^\ast_{\mathbf k}\mathbf
b^{\rm T}_{\mathbf k}\rangle.
\end{equation}
Since $\hbar\langle\mathbf b_{\mathbf k}^\ast\mathbf b_{\mathbf k}^{\rm
T}\rangle$ is diagonal, its elements [the $\hbar n(\mathbf k)$] are
given by the eigenvalues $e_i(\mathbf k)$ of $\boldsymbol\Sigma(\mathbf
k)$. In this way, diagonalisation of $\boldsymbol\Sigma$, a matrix that
can be constructed (via the $\boldsymbol\tau$) entirely from observed
spin displacements in RMC configurations, yields the spin-wave
occupation numbers. The spin-wave frequencies may then be calculated in
a straightforward manner:
\begin{equation}
\omega_i(\mathbf k)=\frac{k_{\rm
B}T}{\hbar}\ln\left[\frac{\hbar}{e_i(\mathbf k)}+1\right].
\end{equation}
By repeating this analysis for a range of wave-vectors, one may
construct a set of spin-wave dispersion curves from the RMC
configurations. The number of $\mathbf k$-points included in these
dispersion curves is limited by the configurational box size used
(which in turn is limited by the available computational resources): a
box representing $n_a, n_b, n_c$ unit cells along axes $\mathbf a,
\mathbf b, \mathbf c$ permits the set of wave-vectors
\begin{equation}
\mathbf k=\frac{i_a}{n_a}\mathbf a^\ast+\frac{i_b}{n_b}\mathbf
b^\ast+\frac{i_c}{n_c}\mathbf c^\ast\qquad i_a, i_b,
i_c\in\mathbb{Z}.
\end{equation}

\begin{figure} \centering
\includegraphics[width=8.0cm]{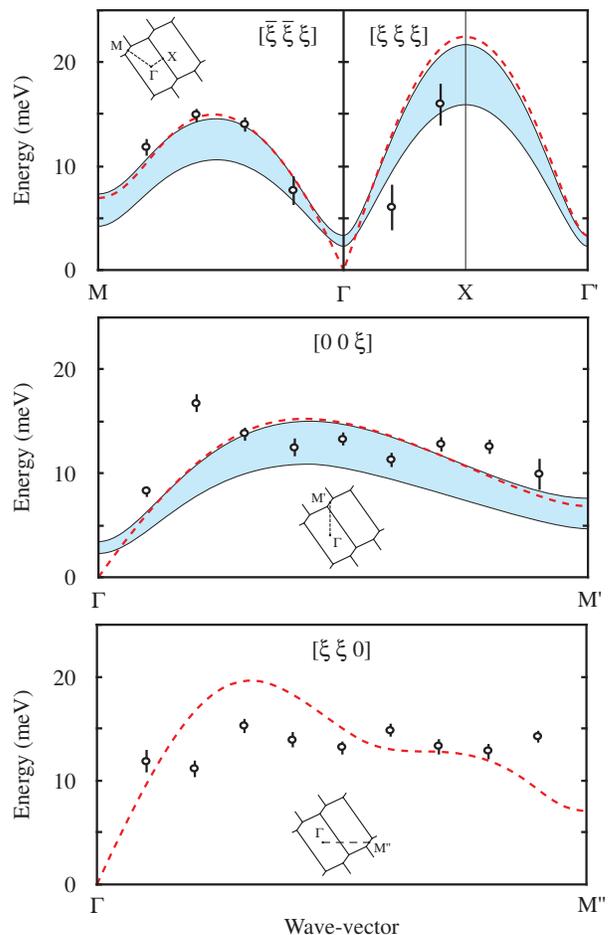}
\caption{\label{fig1}(Colour online) MnO spin-wave dispersion curves
calculated from our RMC configurations (data points). Fits to the INS
data of Pepy~\cite{Pepy_1974} (shaded regions, 88.75\,K$\le
T\le$113.75\,K) and of Kohgi~\cite{Kohgi_1974} (red dashed lines,
$T=100$\,K) are included for comparison. The relevant directions in
reciprocal space are illustrated on a $(1\bar10)$ section of the MnO
reciprocal lattice.}
\end{figure}

We proceeded to apply this analysis to our ensemble of MnO
configurations, calculating spin-wave frequencies across the
$[\bar\xi\bar\xi\xi]$, $[\xi\xi\xi]$, $[00\xi]$ and $[\xi\xi0]$
directions of reciprocal space---these being the axes for which
inelastic neutron scattering (INS) data were available for
comparison~\cite{Pepy_1974,Kohgi_1974}. The $\mathbf k$-point mesh
size (0.1\,r.l.u.) was limited by the number of magnetic---rather
than nuclear---unit cells in the configuration. Calculation at
symmetry-equivalent wave-vectors enabled determination of the errors
involved. While we observed two distinct spin-wave modes at each
wave-vector, inspection of the associated mode displacement vectors
revealed that their relative order in the spin-wave spectrum was not
consistent. This is unsurprising given that the energies of the two
modes are known to be essentially identical at most
wave-vectors~\cite{Pepy_1974,Kohgi_1974,Collins_1973}. Consequently,
the frequencies obtained were also averaged over both branches. Our
results [Fig.~\ref{fig1}]  are compared here to the dispersion curves
given in previous INS studies. Pepy gives data for the
$[\bar\xi\bar\xi\xi]$, $[\xi\xi\xi]$ and $[00\xi]$ branches at
88.75\,K and 113.75\,K, which serve as upper and lower bounds
respectively to the expected energy spectrum~\cite{Pepy_1974}; Kohgi
\emph{et al}. report data at 100\,K, from which they were able to
calculate dispersion curves along the same three branches together
with the $[\xi\xi0]$ direction~\cite{Kohgi_1974}. Collins \emph{et
al.} give data at only at 4.2\,K~\cite{Collins_1973}; as there is
some significant temperature dependence in the dispersion, these
results are not considered further. The curves shown in
Fig.~\ref{fig1} are taken from the spin-interaction models described
in the original reports, refined by fitting to the observed INS
data. The magnon peaks can be relatively broad at such temperatures,
making absolute energies difficult to
assign~\cite{Pepy_1974,Kohgi_1974}. Indeed it is interesting that
the models differ somewhat in their overall energy scale and in the
behaviour of the spin-wave energies near the zone centre.

Our purpose in comparing our calculations with the results of INS
experiments is to establish whether it is at all possible to extract
a similar level of spin-dynamical information from diffraction data.
In this sense, it is promising that the spin-wave energies we
calculate from our RMC configurations agree well with the overall
energy scale observed in INS studies: the level of correlated spin
``motion'' in our RMC configurations is appropriate for the given
temperature and actual spin-wave energies. Moreover, there is some
clear evidence for variation in spin-wave frequency with
wave-vector. Certainly, the form of the $[\bar\xi\bar\xi\xi]$ and
$[\xi\xi\xi]$ branches---and to a lesser extent that of the
$[00\xi]$ branch---follows closely that of the INS-based models. In
addition to the general form of the dispersion curves, one very
encouraging feature is the apparent increase in spin-wave energy at
the M-point relative to the zone centre, which reflects the known
inequivalence of parallel and anti-parallel nearest-neighbour
couplings. However, there are regions where the similarity between
our dispersion curves and those determined using INS is less
precise: our data along $[\xi\xi0]$ show relatively little
dispersion, but do occur over an appropriate energy range. It is not
immediately obvious why the diffraction data should be more
sensitive to some regions of the magnon dispersion than others; what
is known is that modes at different wave-vectors can contribute to
differing extents to the real-space distribution
functions~\cite{Goodwin_2005}.

Despite our employment of state-of-the-art computational resources,
the RMC configurations prepared were not sufficiently large to
determine the precise dispersion behaviour near the BZ origin. The
existence of ``gaps'', their anisotropy and temperature dependence
are often important features of spin-wave spectra, and a method
would have to reflect these to be of general use. However a
limitation of $\mathbf k$-point mesh size is one that can be
expected to abate as computational power increases. Other, more
inherent, limitations will exist. For example, it is likely that
there is some maximum observable spin-wave energy: this is certainly
true for phonon frequencies~\cite{Goodwin_2005}, and magnons suffer
additionally from the restricted $Q$-range over which magnetic
scattering is observable.

At this ``proof of principle'' stage, the important result is that
it is possible to retrieve even broad features of the spin-wave
dispersion from powder diffraction data. Already, one might argue
that relatively sensible parameters for a spin interaction model
could be extracted from dispersion curves such as those in
Fig.~\ref{fig1}, given that some features of the dispersion are well
described, and those that are less-well fixed by the data might
often be constrained by our understanding of spin dynamics.
Moreover, the accuracy of the magnon dispersion curves determined
from diffraction data can be expected to improve significantly. Our
methods of treating the very high quality data one obtains from
instruments such as GEM will improve; access to sufficient
computational power will increase. The potential for powder
diffraction experiments to allow determination of spin-wave
excitation spectra is itself an important result: it provides a
possible mechanism for the exploration of spin dynamics in
newly-characterised materials, for which single crystal samples are
not immediately available.

We acknowledge financial support from EPSRC (U.K.), and from Trinity
College, Cambridge to A.L.G.


\begin{references}
\bibitem{separation_note}
The separation of RMC configurations containing $N$ atoms by at least
$N\ln10$ ``moves'' ensures that the positions or orientations of at
least 90\% of atoms or spins differ from configuration to
configuration.
\end{references}
\end{document}